# Diffusion of indigo molecules inside the palygorskite clay channels

Dejoie C.[a,b,*], Martinetto P.[a], Dooryhée E.[a,c], Brown R.[d], Blanc S.[d], Bordat P.[d], Strobel P.[a], Odier Ph.[a], Porcher F.[e,f], Sanchez del Rio M.[g], Van Eslande E.[h], Walter Ph.[h], Anne M.[a]

[a] Institut Néel, (UPR 2940 CNRS), 25 avenue des Martyrs, BP 166, F-38042 Grenoble Cedex 9, France
[b] Lawrence Berkeley National Laboratory, Advanced Light Source, 1 Cyclotron Road, Berkeley CA 94720, USA
[c] National Synchrotron Light Source-II, Brookhaven, Upton, NY 11973, USA
[d] Institut Pluridisciplinaire de Recherche sur l'Environnement et les Matériaux, CNRS, Hélioparc, 2 avenue Pierre Angot, F-64053 Pau Cedex 9, France
[e] Laboratoire de Cristallographie, Résonnance Magnétique et Modélisation, UHP-CNRS Faculté des Sciences BP 70239, F- 54506 Vandoeuvre-les-Nancy, France
[f] Laboratoire Léon Brillouin, CEA-CNRS, F-91191 Gif-sur-Yvette, France
[g] European Synchrotron Radiation Facility - 6 rue Jules Horowitz, F-38000 Grenoble, France
[h] Centre de Recherche et de Restauration des Musées de France, CNRS, Palais du Louvre, Porte des Lions, 14 Quai François Mitterrand F-75001 Paris, France

* Corresponding author: cdejoie@lbl.gov, catherine.dejoie@grenoble.cnrs.fr

**Abstract**

The search for durable dyes led several past civilizations to develop artificial pigments. Maya Blue (MB), manufactured in pre-Columbian Mesoamerica, is one of the best known examples of an organic-inorganic hybrid material. Its durability is due to the unique association of indigo molecule and palygorskite, a particular fibrous clay occurring in Yucatan. Despite 50 years of sustained interest, the microscopic structure of MB and its relation to the durability remain open questions. Combining new thermogravimetric and synchrotron X-ray diffraction analyses, we show that indigo molecules can diffuse into the channel of the palygorskite during the heating process, replacing zeolitic water and stabilizing the room temperature phases of the clay.

# I. Introduction

Due to the lack of natural stable blue pigments, ancient civilisations were forced to the manufacturing of different synthetic blue pigments. The first known artificial pigment is the Egyptian Blue[1] ($2^{nd}$ millennium BC). Chinese also elaborate a blue/purple pigment found on pottery[2,3] and the terra cotta soldiers of the First Emperor of Qin[4]. Production of Maya Blue in the pre-Colombian MesoAmerica constitutes another example of the know-how of the ancient people[5,6]. In the 60s, the composition of this pigment was elucidated and the durability of Maya Blue was imparted to the association of indigo heat-treated with fibrous clays (palygorskite and sepiolite)[7]. The exceptional chemical stability of Maya Blue is shown by the persistence of the blue colour in acid and oxidising conditions[8,9]. Maya Blue is now considered as one of the first ever artificial organic-inorganic hybrid[10], and its exceptional stability is a source of inspiration in the design of new hybrid material[11,12,13,14,15].

Palygorskite, main ingredient of Maya Blue, belongs to the phyllosilicate family. This fibrous clay presents parallel channels running along the fibres in the c direction (Fig. 1a) with an opening cross section of 6.4*3.7 Å². At room temperature, tunnels are filled with weakly bound zeolitic water, released by moderate heating or under vacuum. This opens the way to the replacement of water molecule by gaseous or organic molecules[16,17,18,19,20,21,22]. Palygorskite of general formula $Si_8O_{20}Al_2Mg_2(OH)_2(H_2O)_4 \cdot 4H_2O$ is substituted by Al (and Fe) atoms. The Mexican palygorskite used in the formation of Maya Blue is composed of a mixture of monoclinic and orthorhombic polymorphs[23]. Indigo (Fig. 1b), used by ancient Maya to form archaeological Maya Blue, was well known by ancient civilizations[24]. Indigo is composed of indigotin $C_{16}H_{10}N_2O_2$ (3H-indol-3-one, 2-(1,3-dihydro-3-oxo-2H-indol-2-ylidene)-1,2-dihydro), a quasi-planar molecule of approximate dimensions 5*12 Å². The deep blue colour of the dye is attributed to the presence of the NH donor groups and the C=O acceptor groups substituting the central double bond C=C of the molecule[25].

The reasons for the colour durability of the Maya Blue pigment are the object of an on-going discussion, and several articles recently discussed the nature and the location of the clay-tied indigoid entity. According to Van Olphen[7] and Chiari et al.[26], the indigo molecule preferentially lies in the grooves of the surface of the palygorskite particles. On the other hand, Kleber et al.[27], Chiari et al.[23], and Fois et al.[28] assumed that indigo substitutes with internal zeolitic water of palygorskite and enters the channels. A third model is proposed by Hubbard et al.[29], in which indigo does not enter the channels of palygorskite and stays at the entrance of the clay, preventing further water departure. A moderate thermal treatment, required to obtain the stable dye, is said to activate organic-inorganic interaction, inducing a lightning color shift, from dark blue to turquoise[30]. Hydrogen bonds have been found to settle between the amine and carbonyl groups of indigo and structural water of clays[23,**Erreur ! Signet non défini.**,31]. Transformation of indigo into an intermediate oxidized form, the dehydroindigo, has also been introduced[32,42] with possible bonding between nitrogen of the organic dye and aluminum site of palygorskite[33].

This study focalizes onto the dehydration/rehydration process and the structural evolution of the palygorskite clay in the absence and the presence of the indigo molecule. We investigate effects of the thermal treatment during the synthesis of Maya Blue reproductions using a combination of thermogravimetric (TGA), calorimetric (DSC) analyses, and synchrotron X-ray powder diffraction (XRD). We show that indigo induces a stabilization of the room temperature phases of the palygorskite. Our study also demonstrates the complexity of the diffusion process of the indigo molecules inside the channels of the clay, which is really sensitive to the synthesis conditions (initial concentration and heating process).

## II. Experimental

Palygorskite clay (also called attapulgite) came from Ticul in the Yucatán peninsula (Mexico), a major source of clay for the manufacturing of Maya Blue for more than eight centuries. Synthetic indigo (Sigma-Aldrich) was used in the formation of all the hybrid samples. Clay and indigo were hand-ground and mixed in a mortar. Before XRD experiments, the powders were hand-packed in 1mm capillary glass tubes. To avoid any re-hydration, samples were directly heated in open capillaries for 5 hours at 200°C and 250°C, and sealed before going back to room temperature. "Ex situ" samples were pressed into a pellet to enhance contact between the two components. The pellets were baked in air in an oven for different duration time and different temperatures. After the heating phase, pellets were re-ground and washed with acetone to remove excess indigo, and exposed several days in air at room temperature to perform rehydration process. Chemical stability of the samples was tested in nitric acid condition. A few milligrams of the powder was stirred for 10 min in concentrated $HNO_3$ (14M). Stability of the sample was attested by the persistence of the blue colour.

During the X-Ray Diffraction measurements (XRD), capillary glass tubes were mounted on the goniometer head of the 7-circles diffractometer of beamline BM02-ESRF[34]. Powder diffraction data were collected in the high-resolution 2θ step scanning mode at room temperature, and data were processed using the Fullprof software[35]. Thermogravimetric (TGA) and calorimetric (DSC) analyses were conducted on a SETARAM TAG 24 thermoanalyser instrument (Institut Néel Grenoble). Samples (~15 mg) were heated from 25 to 800 °C at a heating rate of 10°C/min under flowing nitrogen.

## III. Results and discussion

### III.1. Palygorskite clay

TGA and DTA curves for the palygorskite clay are shown on Fig. 2. At room temperature, three types of water molecules are found in palygorskite: weakly-bond surface water, zeolitic water inside the channels, and structural water molecules tightly bond, completing cations (Mg or Al) coordinance at the border of octahedral sheets. A first weight loss of 9.4% between 25°C and 150°C is attributed to the departure of surface water (endothermic peak at 90°C) and zeolitic water (second endothermic peak at 130°C). This temperature range for zeolitic water departure is in good agreement with previous works[36,37,38,39]. Structural water loss is decomposed into two steps. The first half of bounded water molecules is lost between 200°C and 300°C (4% on the TGA curve, with the corresponding endothermic peak at 250°C). Second part of structural water (5.6%) is lost between 300°C and 550°C. The associated endothermic peak is found around 470°C. These two structural water loss are consistent with literature[40,36,37,38].

Refinement results (full pattern matching) for palygorskite for different heating temperature and before rehydration are summarized in Table 1. From 25°C to 200°C, a decrease of the cell volume for the monoclinic (C2/m) and the orthorhombic (Pbmn) phases is observed (Table 1), mainly due to the decrease of the a parameter. Post et al.[40] already report this structural evolution related to the departure of zeolitic water, but only for a purely monoclinic palygorskite. Departure of the first half of structural water (see TGA data) causes a structural change and the emergence of a new monoclinic phase ($P2_1/a$). This phase transition is attributed to the folding of the structure by rotation of the phyllosilicate ribbons about an axis through the Si-O-Si corner bonds that link the ribbons[40,41,40]. At 250°C, cell volume for the two primary phases of palygorskite (monoclinic C2/m and orthorhombic Pbmn) is severely decreased compared with the ones refined at 25°C and 200°C, as the result of the a parameter evolution. A higher temperature (~300°C) is required for a complete phase transition (not shown).

**III.2. Indigo@palygorskite hybrids**

In order to follow the water departure process during the heating phase leading to the formation of MB reproductions, we performed *in situ* thermogravimetric experiments. A mixture of raw clay and indigo powder (10%wt.) was heated from 20°C to 190°C, directly inside the TGA furnace. As indigo is known to be stable up to 350-400°C, the weight loss measured at 150°C on the TGA curves is directly related to the departure of surface and zeolitic water. The weight loss measured at 150°C for the *in situ* sample is comparable to that of the raw palygorskite (Fig. 3a), meaning that both surface and zeolitic waters are able to depart the clay during the hybrid formation process. This result was compared to a 10%wt. *ex situ* sample baked 5 hours at 190°C, and able to rehydrate after cooling back to room temperature. The weight loss at 150°C is decreased compared to that of the *in situ* sample. We conclude that presence of indigo prevents complete rehydration of the clay. Corresponding DSC curves for the *in situ* and *ex situ* samples are shown on Fig. 3b. The endothermic peak related to surface water is not really affected, contrary to the one related to zeolitic water which is clearly decreased for the *ex situ* sample. During the heating process, indigo diffuses into the channels after dehydration of the clay and then prevent part of zeolitic water to come back when cooling back to room temperature.

We then performed new thermogravimetric experiments on different *ex situ* MB reproductions, to look at the rehydration process of the clay in presence of indigo molecules. Fig. 4 represents the sum of surface and zeolitic water content measured at 150°C on the TGA curves, depending on the initial indigo concentration, temperature formation and duration of the heating phase. For a 10%wt. concentration of indigo and constant 5h duration time, rehydration of the clay decreases when temperature formation increase. For samples obtained at 190°C (5h duration time), we observe that rehydration decreases when initial concentration of indigo increases. The effect of the duration time of the heating process was checked at 150°C for a 10%wt. concentration. In this case, rehydration was found to decrease with the increase of the duration time. We underline here all the complexity of the diffusion process of indigo molecules in the palygorskite channels. The TGA results show strong dependence on the pigment production conditions such as, the indigo concentration, the temperature reached, and the duration of the heating process. Different sample preparation methods can lead to significant differences in surface and zeolitic water contents. This observation can contribute to the different results and models found in the literature for the indigo location in the Maya Blue pigment.

XRD experiments are also performed on non-rehydrated indigo@palygorskite samples previously heated to check any structural evolution of the clay in presence of indigo. Full pattern matching results are summarized in Table 2. Cell parameters measured at 200°C and 250°C for the monoclinic and the orthorhombic phases of the indigo@palygorskite sample are similar to the ones found for the raw palygorskite at 25°C (Table 1). The decrease of the *a* parameter at higher temperature is no longer observed in presence of 10%wt. of indigo. Moreover, contrary to the raw palygorskite at the same temperature, the folded phase is not present at 250°C for the indigo@palygorskite hybrid (Fig. 5). This observation strongly suggests that the folding of the clay is hindered in the presence of indigo. Such a prevention of folding due to the presence of indigo molecules into clay channels for an indigo@sepiolite hybrid[42] was reported by Ovarlez *et al*. To our knowledge, this is the first time that this aspect is also shown for palygorskite. The long range structure of the palygorskite clay is thus affected by the presence of indigo, which is in agreement with the presence of the organic dye inside the channels of the clay.

The evolution of the *a* parameter of the palygorskite monoclinic phase as a function of the temperature and for different indigo concentrations is given on Fig. 6. Same evolution is obtained using the *a* parameter of the orthorhombic phase (not shown). The *a* parameter evolution appears to be dependent on the indigo concentration: it decreases with temperature as the concentration decreases. With a 2%wt. concentration, the folded phase is present at 250°C (not shown). With a 10%wt concentration, the diffraction pattern of the hybrid is insensitive to temperature up to 300 °C, with no folded phase emergence. These XRD results showing the dependence of structure change as a function of indigo content substantiate the TGA results. These results also support that the substitution of zeolitic water by indigo has strong dependence on the pigment synthesis process (concentration of indigo and heating phase).

## IV. Conclusion

This study of modern replication of archaeological Maya Blue focuses on the effect of the diffusion process of indigo under heating on the palygorskite clay (structure change and water content). Diffusion of indigo molecules in the channels prevents the clay from folding at high temperatures. The pigment maintains the XRD signature of the room temperature phases of palygorskite. As demonstrated by TGA data, water reoccupies remaining free channel volume on cooling. Rehydration depends on the synthesis conditions, such as initial indigo concentration, temperature formation, and duration of the heating phase. It is important to point out that the commonly adopted content of indigo (<2%wt.) for the archeological Maya Blue, which is responsible for the light blue-turquoise hue, leads to weak signals in the TGA and the XRD experiments comparing to the raw clay content. Such weak signals hinder us to fully understand the formation process of ancient Maya Blue and the microscopic structure of MB and its relation to the durability. In this study, Maya Blue was obtained by using high indigo content (10%wt.) in modern reproduction, which gives us a clear picture of the indigo@palygorskite hybrid formation.

**Acknowledgements**

Synchrotron diffraction measurements at the ESRF-BM02 beamline benefited from the support of J-F. Bérar, N Boudet, S. Arnaud and B. Caillot. C.D. acknowledges CIBLE PhD and MACODEV grants from Région Rhône-Alpes.

**Table caption**

**Table 1 - Crystallographic results of the full pattern fitting of raw palygorskite at 25°C, and after heating 5 hours at 200°C and 250°C.**

| T (°C) | 25°C | | 200°C | | 250°C | | |
|---|---|---|---|---|---|---|---|
| Symmetry | Monoclinic | Orthorhombic | Monoclinic | Orthorhombic | Monoclinic | Orthorhombic | Monoclinic |
| Space group | C2/m | Pbmn | C2/m | Pbmn | C2/m | Pbmn | $P2_1/a$ |
| a (Å) | 13.282(4) | 12.790(3) | 13.105(8) | 12.667(10) | 12.445(9) | 12.557(6) | 10.946(13) |
| b (Å) | 17.798(3) | 17.845(4) | 17.889(14) | 17.879(16) | 17.837(9) | 18.079(6) | 15.279(19) |
| c (Å) | 5.250(1) | 5.225(1) | 5.250(4) | 5.217(3) | 5.177(2) | 5.206(2) | 5.265(5) |
| α (°) | 90 | 90 | 90 | 90 | 90 | 90 | 90 |
| β (°) | 107.60(2) | 90 | 107.68(9) | 90 | 107.89(7) | 90 | 96.49(9) |
| γ (°) | 90 | 90 | 90 | 90 | 90 | 90 | 90 |
| Volume (Å$^3$) | 1183.1(5) | 1192.4(5) | 1173(1) | 1182(2) | 1093(1) | 1181.7(8) | 874(2) |
| Rp (%) | 7.50 | | 12.6 | | 15.1 | | |
| Rwp (%) | 9.64 | | 16.5 | | 16.3 | | |
| Re (%) | 2.6 | | 7.7 | | 9.9 | | |
| Chi2 | 13.4 | | 4.6 | | 2.7 | | |

**Table 2 - Crystallographic results of the full pattern fitting of a 10%wt. indigo@palygorskite sample, heated 5 hours at 200°C and 250°C.**

| T (°C) | 200°C | | 250°C | |
|---|---|---|---|---|
| Symmetry | Monoclinic | Orthorhombic | Monoclinic | Orthorhombic |
| Space group | C2/m | Pbmn | C2/m | Pbmn |
| a (Å) | 13.284(7) | 12.766(6) | 13.247(9) | 12.766(7) |
| b (Å) | 17.868(10) | 17.883(10) | 17.881(14) | 17.888(14) |
| c (Å) | 5.247(2) | 5.236(2) | 5.264(3) | 5.247(2) |
| α (°) | 90 | 90 | 90 | 90 |
| β (°) | 107.53(6) | 90 | 107.63(9) | 90 |
| γ (°) | 90 | 90 | 90 | 90 |
| Volume (Å$^3$) | 1187(1) | 1195(1) | 1188(1) | 1198(1) |
| Rp (%) | 11.0 | | 11.7 | |
| Rwp (%) | 13.5 | | 14.6 | |
| Re (%) | 5.5 | | 6.7 | |
| Chi2 | 6.1 | | 4.8 | |

**Figure caption**

**Figure 1 - a) Structure of monoclinic palygorskite ((001) direction)[40] — b) Indigo molecule.**

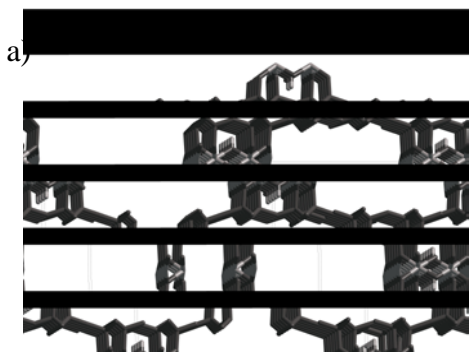

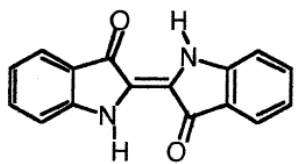

**Figure 2 -  TGA and DSC curves for the raw palygorskite.**

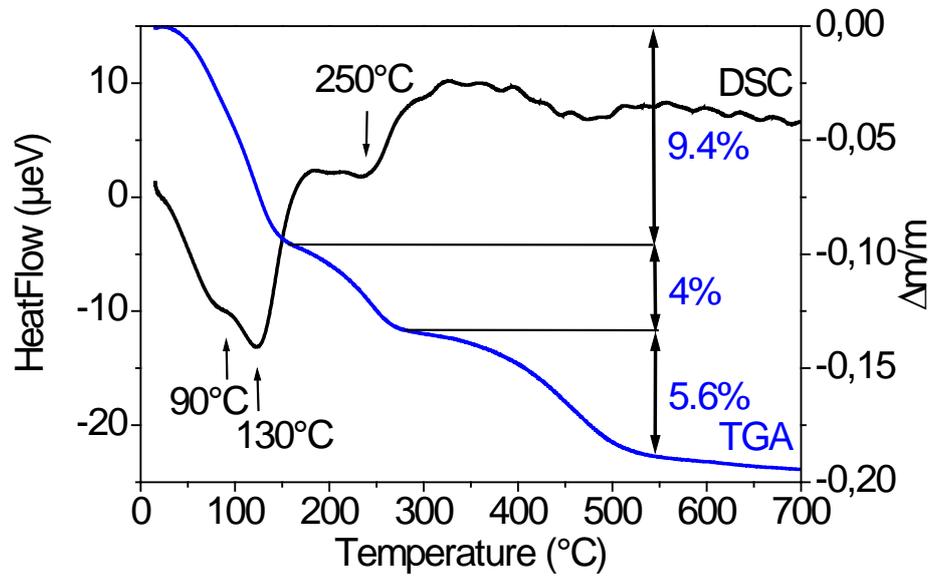

**Figure 3** – a) Thermogravimetric experiments for palygorskite clay (□), *in situ* 10%wt. indigo/clays mixture (○), and previously heated *ex situ* 10%wt. indigo@clay sample (Δ) - b) Calorimetric experiments for *in situ* 10%wt. indigo/clays mixture (○), and previously heated *ex situ* 10%wt. indigo@clay sample (Δ).

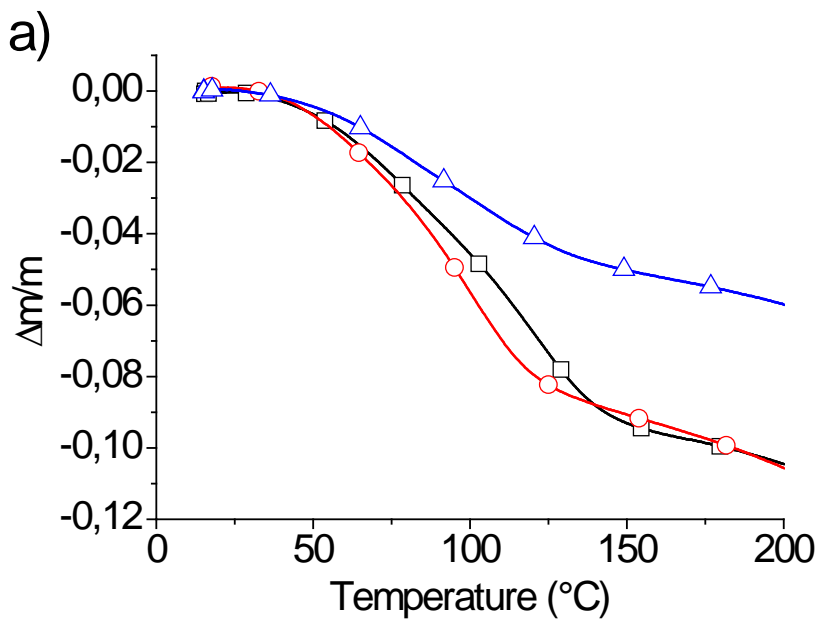

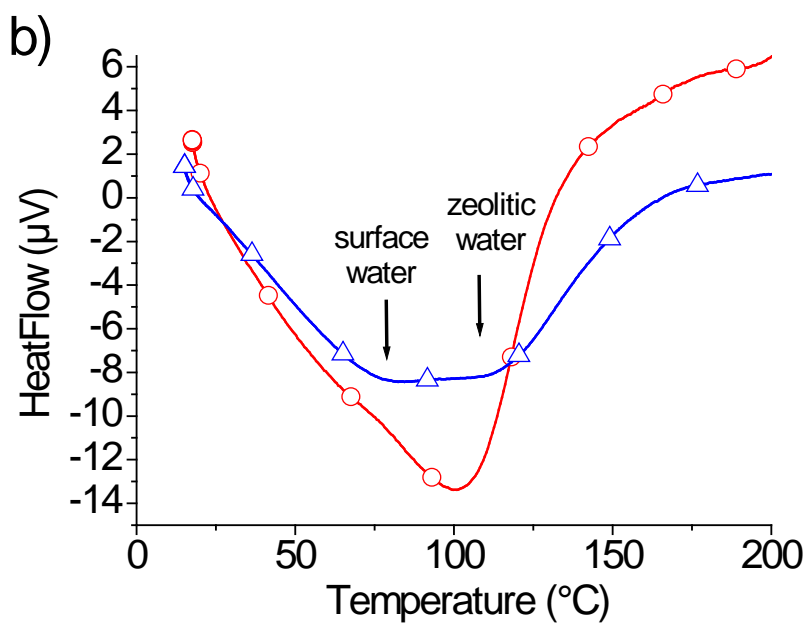

**Figure 4 – Water loss departure measured at 150°C of different *ex situ* indigo@clay samples. Non marked (Δ) samples are 10%wt., heated 5 hours. 2%wt. and 5%wt. samples were heated 5 hours at 190°C. Samples at 150°C heated for different duration time were 10%wt. Standard deviations of each point are within the plotting symbols.**

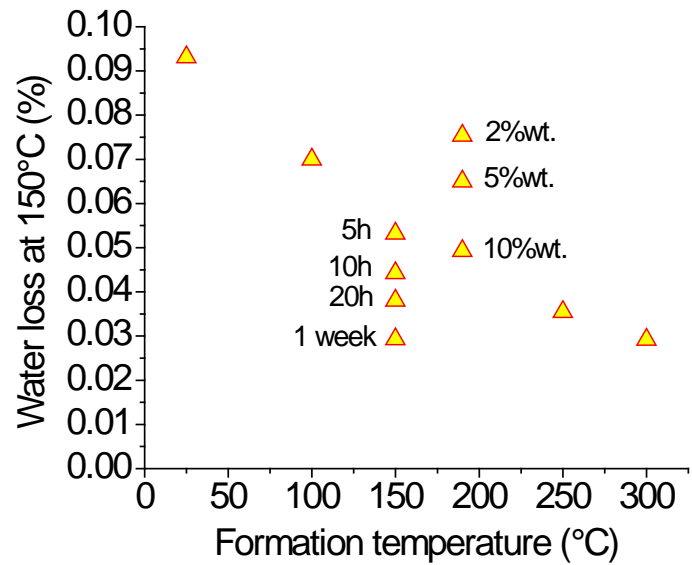

**Figure 5** - Diffraction diagrams of a) raw palygorskite heated 5 hours at 250°C, and b) a 10%wt. indigo/palygorskite sample heated 5 hours at 250°C (red points: experimental data ; black line: calculated diagram ; green: Bragg positions ; blue bottom line: residuals). Three phases are present in the raw palygorskite diagram, due to the emergence of the folded one.

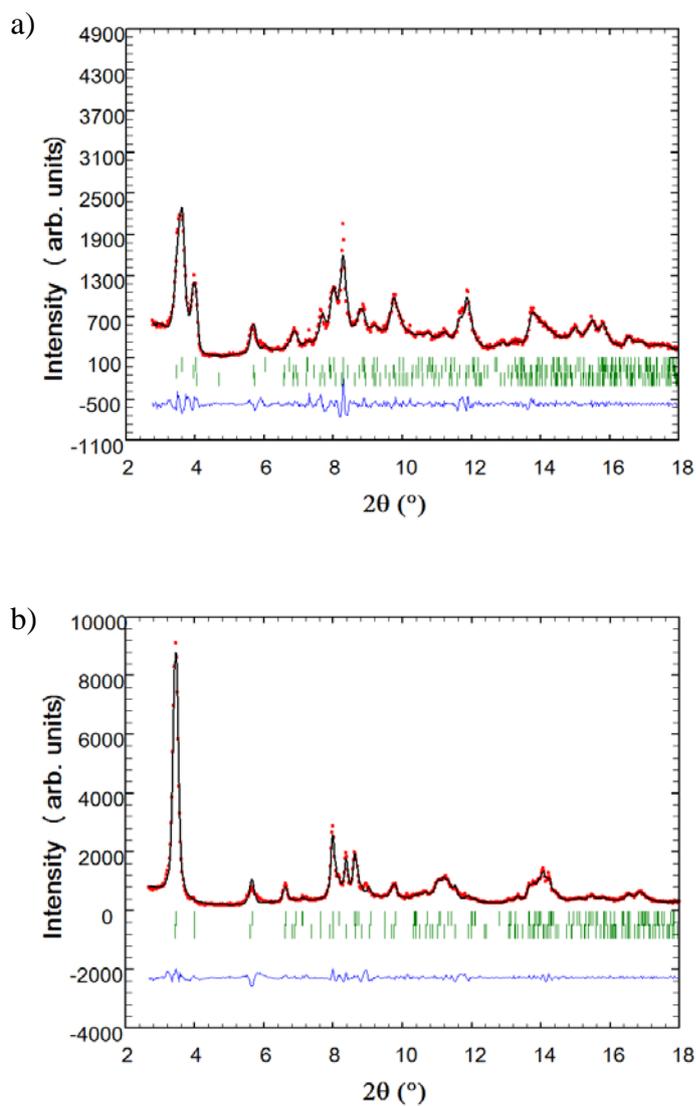

**Figure 6** - Evolution of the *a* parameter of palygorskite (monoclinic phase) *vs.* temperature and indigo concentration. The errors determined by the refinement are within the plotting symbols.

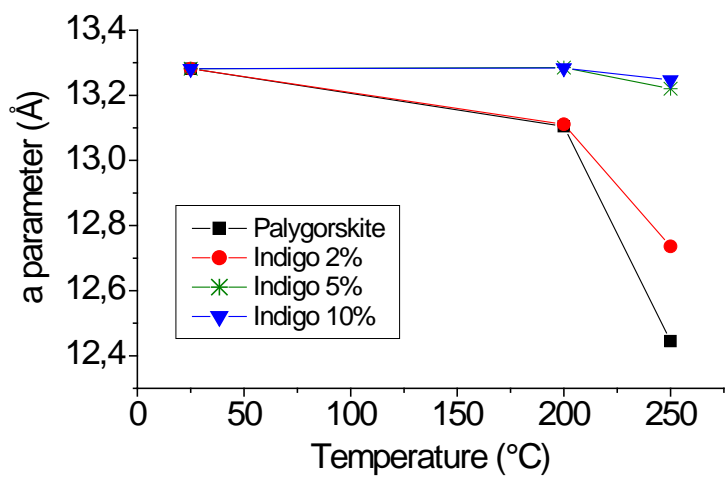

[30] D. Reinen, P. Kohl and C. Muller: The nature of the colour centres in "Maya Blue" – the incorporation of organic pigment molecules into the palygorskite lattice. Z. Anorg. Allg. Chem. 630, 97 (2004).

[31] R. Giustetto, D. Levy and G. Chiari: Crystal structure refinement of Maya Blue pigment prepared with deuterated indigo, using neutron powder diffraction. Eur. J. Mineral. 18, 629 (2006).

[32] A. Domenech, M.T. Domenech-Carbo and M.L. Vazquez de Agredos Pascual : Dehydroindigo: A New Piece into the Maya Blue Puzzle from the Voltammetry of Microparticles Approach. J. Phys. Chem. B. 110, 6027 (2006).

[33] F.S. Manciu, L. Reza, L.A. Polette, B. Torres and R.R. Chianelli: Raman and infrared studies of synthetic Maya pigments as a function of heating time and dye concentration. J. Raman Spectrosc. 39, 1257 (2008).

[34] http://www.esrf.eu/UsersAndScience/Experiments/CRG/BM02

[35] Rodriguez-Carvajal: Recent Advances in Magnetic Structure Determination by Neutron Powder Diffraction. Physica B. 192, 55 (1993).

[36] H. Hayashi, R. Otsuka and N. Imai: Infrared study of sepiolite and palygorskite on heating. Am. Mineral. 53, 1613 (1969).

[37] G. Artioli and E. Galli: The crystal structures of orthorhombic and monoclinic palygorskite. Mater. Sci. Forum. 166, 647 (1994).

[38] G. Artioli, E. Galli, E. Burattini, G. Cappuccio and S. Simeoni: Palygorskite from Bolca, Italy: a characterization by high-resolution synchrotron radiation powder diffraction and computer modelling. N. Jb. Miner. Mh. 5, 217 (1994).

[39] R. Giustetto, F.X. Llabres I Xamena, G. Ricchiardi, S. Bordiga, A. Damin, R. Gobetto and M.R. Chierotti: Maya Blue : a computational and spectroscopic study. J. Phys. Chem. B. 109, 19360 (2005).

[40] J.E. Post and P.J. Heaney: Synchrotron powder X-ray diffraction study of the structure and dehydration behaviour of palygorskite. Am. Mineral. 93, 667 (2008).

[41] A. Preisinger: Sepiolite and related compounds: its stability and applications. Clays Clay Miner. 10, 365 (1963).

[42] S. Ovarlez, F. Giulieri, A.M. Chaze, F. Delamare, J. Raya and J. Hirschinger: The incorporation of indigo molecules in sepiolite tunnels. Chem. Eur. J. 15, 11326 (2009).